\documentclass{article}

\usepackage{amsmath,amsfonts,amssymb,amstext,verbatim,color,times,epsfig,graphicx}

\usepackage{amsbsy}
\usepackage{subcaption}
\usepackage{booktabs,siunitx}
\usepackage{caption}

\usepackage{lettrine}
\usepackage{xcolor}
\usepackage{hyperref}
\hypersetup{breaklinks=true}

\usepackage{arxiv}

\usepackage[utf8]{inputenc} 
\usepackage[T1]{fontenc}    
\usepackage{hyperref}       
\usepackage{url}            
\usepackage{booktabs}       
\usepackage{amsfonts}       
\usepackage{nicefrac}       
\usepackage{microtype}      
\usepackage{lipsum}
\usepackage{graphicx}
\graphicspath{{media/}}     

 \title{Vertical stabilization of tokamak plasmas\\ via Extremum Seeking}

\author{
  L.~E.~di~Grazia \\
  Consorzio CREATE \\
  Dipartimento di Ingegneria, Universit\`a degli Studi della Campania ``L.~Vanvitelli" \\
  via Roma 29, Aversa (CE), 80131, Italy \\
  \texttt{luigiemanuel.digrazia@unicampania.it} \\
  \And
  S.~Dubbioso, G.~De~Tommasi, M.~Mattei, A.~Pironti\\
  Consorzio CREATE \\
  Dipartimento di Ingegneria Elettrica e delle Tecnologie dell'Informazione, Universit\`a degli Studi di Napoli Federico II \\
  via Claudio 21, 80125, Napoli, Italy \\
  \texttt{sara.dubbioso@unina.it, detommas@unina.it, pironti@unina.it}  \\
  \And
  A.~Mele \\ 
  Dipartimento di Economia, Ingegneria, Società e Impresa, Universit\`a degli Studi della Tuscia  \\
  Campus Riello - Blocco F, largo dell'Università, 01100, Viterbo, Italy\\
  \texttt{adriano.mele@unitus.it} 
}

\begin{document}
\maketitle

\begin{abstract}                          
In this paper we propose a vertical stabilization~(VS) control system for tokamak plasmas based on the extremum seeking (ES) algorithm. 
The gist of the proposed strategy is to inject an oscillating term in the control action and exploit a modified ES algorithm in order to bring to zero the \emph{average} motion of the plasma along the unstable mode. In this way, the stabilization of the unstable vertical dynamic of the plasma is achieved. 
The approach is validated by means of both linear and nonlinear simulations of the overall~ITER tokamak magnetic control system, with the aim of demonstrating robust operation throughout the flat-top phase of a discharge and the capability of reacting to a variety of disturbances.
\end{abstract}


\section{INTRODUCTION}\label{section:Introduction}


Nuclear fusion is foreseen as a possible source of energy for the next century~\cite{EUROfusionRM,USACR}. A big effort has been made, since the end of World War~II, with the aim of developing its peaceful use toward the realization of a power plant. The tokamak concept arose in the~50s-60s of the last century, as one of the most promising experimental devices, 

5 warnings
 aimed at proving the feasibility of energy production by means of nuclear fusion on Earth~\cite{wesson2011tokamaks}. 

Since the mid~70s, many international projects have been successfully established to build and operate tokamaks all around the world. The~JET tokamak~\cite{ScienceJET} is still the world's largest, although it will be soon exceeded in size by the joint~EU-Japan project~JT-60SA~\cite{barabaschi2019progress} and by~ITER~\cite{ITER}, which is an international enterprise involving~EU, India, People's Republic of China, Russia, South Korea and~ USA, and which is currently under construction in France.

\begin{figure}[!thb]
\begin{center}
\includegraphics[width=11cm]{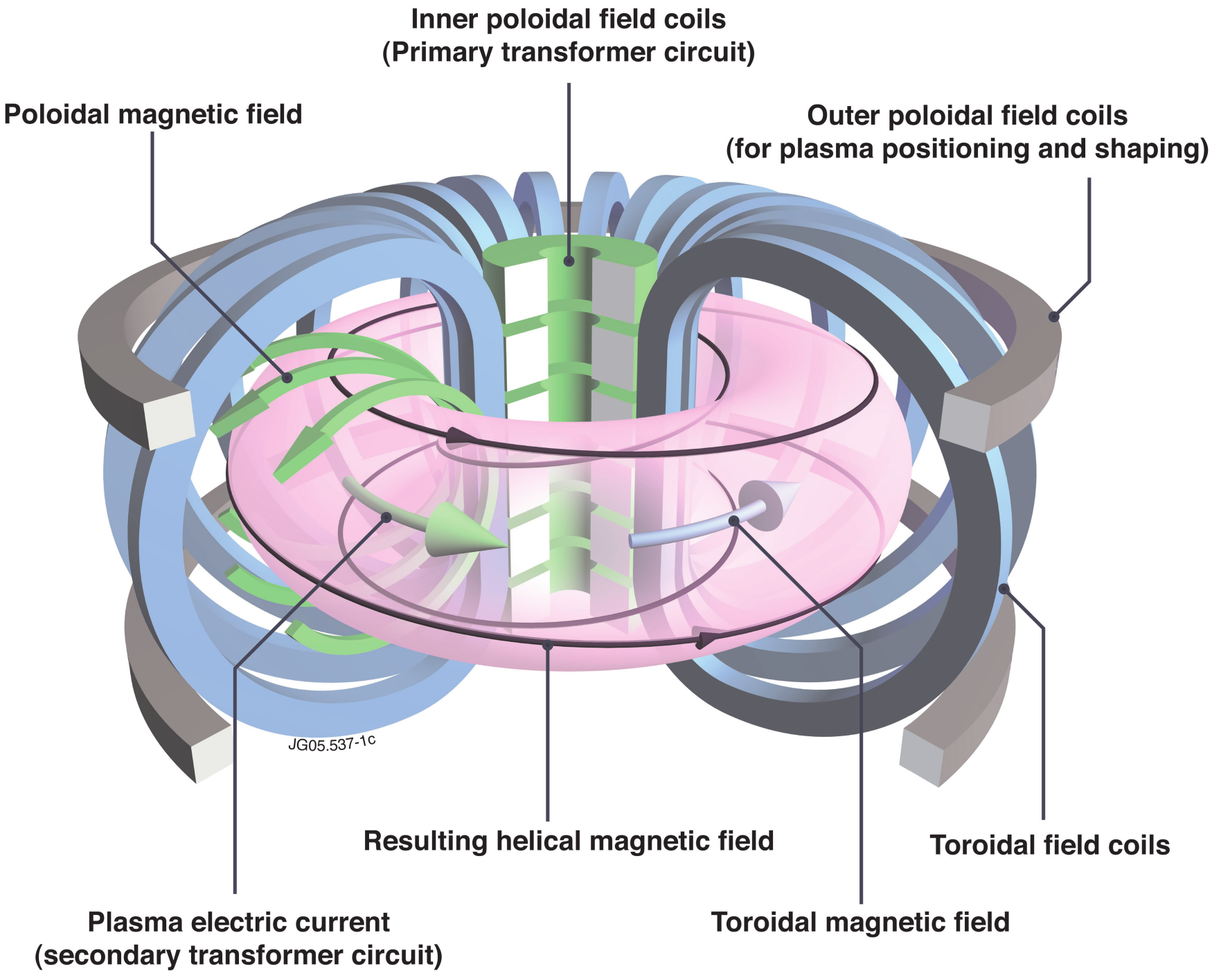}
\caption{Simplified scheme of a tokamak fusion device.}\label{figure:tokamak}
\end{center}
\end{figure}

\medskip 
In a tokamak, a fully ionized gas of hydrogen ions, called \emph{plasma}, is confined by magnetic fields and heated to temperatures of tens to hundreds millions degrees. At such high temperatures, collisions between ions can overcome the Coulomb repulsive forces, resulting into fusion reactions. Plasma confinement is achieved by means of both toroidal and poloidal magnetic field sources. In particular, the toroidal field component is produced by a set of coils wrapped around the vacuum vessel (see the blue coils in Figure~\ref{figure:tokamak}), while the poloidal one is generated by the presence of a plasma current induced in the ionized gas, and by a set of toroidally continuous coils (in grey in Figure~\ref{figure:tokamak}), called Poloidal Field (PF) coils.

\begin{figure}[!thb]
\begin{center}
\includegraphics[width=11cm]{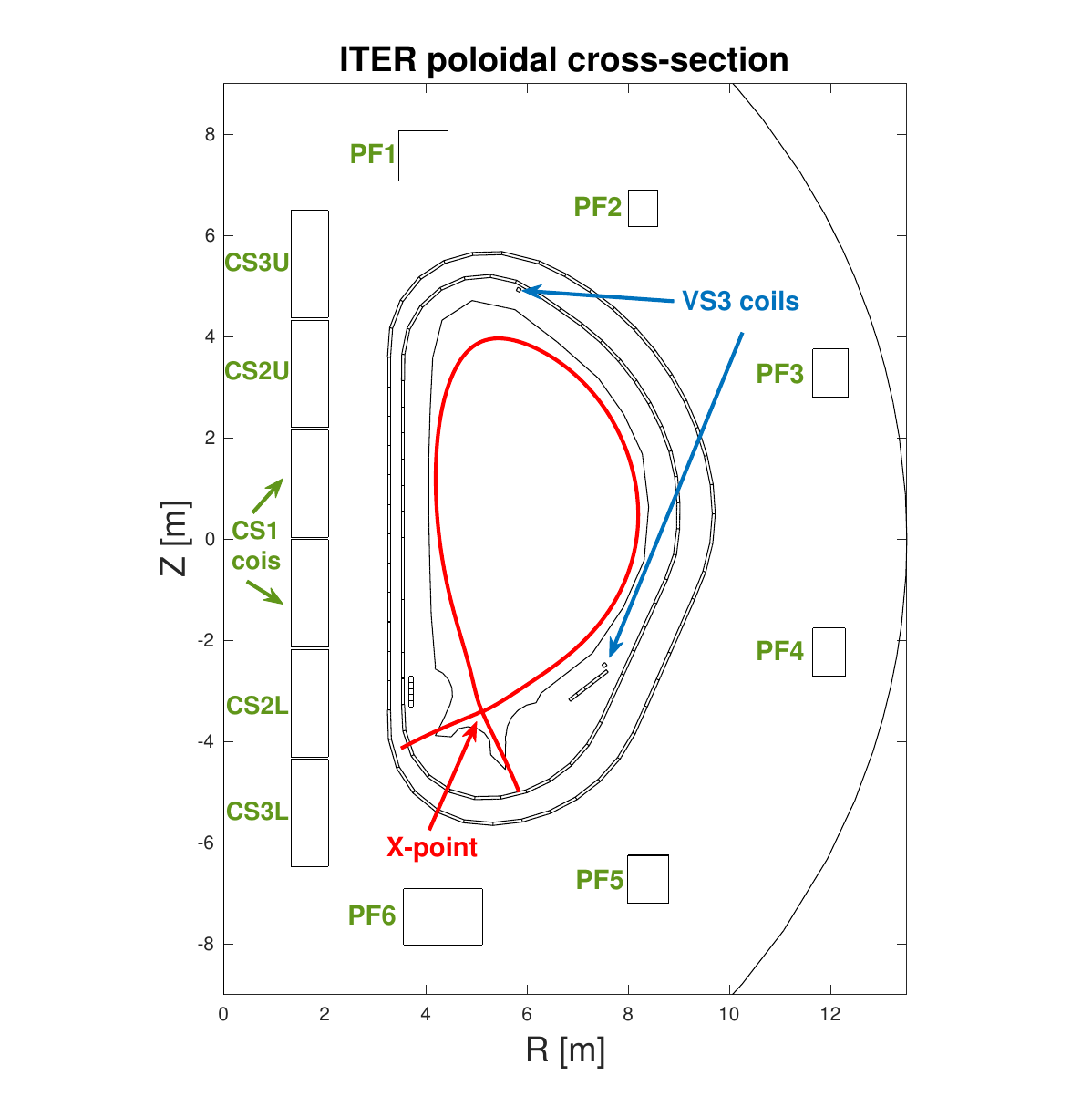}
\caption{Poloidal cross-section of an elongated~ITER tokamak plasma. The red curve shows the boundary of an elongated plasma. The coils of the superconductive~PF circuits are labeled in green, while the blue labeled in-vessel coils are the ones that form the~$VS3$ circuit, which is the actuator for the~ITER Vertical Stabilization system.}\label{figure:ITER_pcs}
\end{center}
\end{figure}

Operation of large tokamaks such as~ITER calls for the solution of several challenging control problems, among which there is the so called \emph{magnetic control problem}, i.e.~the control of the current induced into the plasma, as well as of its shape and position, by regulating the poloidal field produced by the currents flowing in the~PF coils~\cite{AriolaPironti:Springer}. Accurate plasma position and shape control is needed for several reasons: from the avoidance of wall interactions~\cite{de2014shape} to the optimization of divertor pumping~\cite{calabro2015east}. In a tokamak, the plasma position and shape control task is further complicated by the fact the plasma exhibits a vertical instability, due to the elongated shapes typically pursued~\cite{walker2009feedback} (see the elongated cross-section of the~ITER plasma reported in Figure~\ref{figure:ITER_pcs}).

\medskip
Different model-based approaches have been proposed in literature to solve the vertical stabilization problem in a robust fashion, including nonlinear adaptive control~\cite{scibile2001discrete}, MPC~\cite{gerkvsivc2013vertical} and multi-objective optimization techniques~\cite{de2017robust}. In many cases the adopted control approach has been tailored taking into account the features of the specific experimental device: see~\cite{Sartori:JET} for the vertical stabilization of the~JET tokamak, \cite{schuster2005plasma} for the~DIII-D system, and~\cite{CST:2010} for a model-based~ITER~VS system that has been also tested on the~EAST tokamak~\cite{albanese2017iter,de2017plasma}. 

Despite the different proposed solutions, the performance of any existing~VS system strongly depends on the \emph{growth rate}~$\gamma$ of the instability, usually defined as the unstable eigenvalue of the linearized plasma response model obtained around the considered configuration. The eigenvector associated to $\gamma$ describes the behaviour of the plasma and of the currents in the passive structures along the unstable direction. A possible~VS control approach could be to adapt the control gains as function of~$\gamma$. However, the estimation of the unstable eigenvalue is based on the real-time reconstruction of the plasma equilibrium~\cite{bao2020real}, which is still a computationally demanding task, if compared with the time scale the~VS system should react in. 

\medskip
In this paper we introduce a plasma vertical stabilization system that exploits the ES-based approach originally proposed in~\cite{scheinker2017model}. In particular, in their work Scheinker and Kristi\'c have shown that it is possible to stabilize \emph{on average} an unstable plant by minimizing a properly chosen candidate Lyapunov function of the plant state with a suitable choice of the control action. 
Starting from these theoretical results, here we show that it is possible to achieve a satisfactory level of robustness for the vertical stabilization system during the overall flat-top of an~ITER discharge by exploiting the \emph{model agnosticity} of the~ES, i.e. without the necessity of a detailed plasma model. Indeed, we show that only a single linear model of the plasma at the flat-top is sufficient to design a~VS algorithm that works for the entire flat-top, despite the variation of plasma shape and current density distribution. As it will be discussed in what follows, this unique linear model is required to design a reduced order Kalman filter which estimates the plasma motion along the unstable mode. 

The validation of the proposed~ES-based vertical stabilization approach and the assessment of its robustness are carried out by both linear and nonlinear simulations. By means of the linear simulations we prove that the proposed approach robustly stabilizes a broad family of different plasma models, although the embedded Kalman filter is always the same. On the other hand, nonlinear simulations that involve the solution of a free boundary evolutionary problem are used to show the robustness of the~ES-based VS throughout the overall~ITER discharge. 

It is worth to remark that the considered simulation setup includes the complete~ITER magnetic control system, i.e.~it also takes into account the interaction of the proposed~ES-based VS with the plasma current and shape controllers. Moreover, the~ES-based approach does not necessarily require the use of a linear power amplifier for the~PF circuit used for~VS. Therefore, in the presented simulations we consider a model of a switching power supply, which provides a faster time response with respect to linear amplifiers. 



The proposed approach can be regarded as an \emph{almost} model-free one. Although there is an interest in developing model-free plasma control techniques, to the best of the authors knowledge data-driven vertical stabilization approaches are still missing in the literature, even though this kind of techniques have been proposed for tracking problems, such as the control of plasma internal profiles~\cite{shi2017data,wakatsuki2019safety}.

\medskip

The rest of the paper is structured as follows: Section~\ref{section:NonlinearSim} introduces the modelling and simulation environment used to perform both the validation and the performance assessment of the proposed stabilization approach.
The magnetic control problem in tokamak fusion devices and a possible architecture for the~ITER plasma magnetic control system are presented in Section~\ref{section:MagneticControlProblem}, and the vertical stabilization problem, with a particular focus on~ITER, is discussed in Section~\ref{subsection:VSproblem}.
Then, the proposed~ES-based control algorithm to solve such stabilization problem is described in Section~\ref{section:VS_ES}, while the simulation results are discussed in Section~\ref{section:SimResults}. In particular, Section~\ref{subsection:LinearSim} describes the simulations performed over a family of different linearized models, while in Section~\ref{nonlinearsimulations} a set of nonlinear simulations is described, with the aim of validating the proposed approach.
Eventually, some conclusive remarks are given.

\section{NONLINEAR MODELLING OF PLASMA/CIRCUIT DYNAMICS}\label{section:NonlinearSim}
Mathematical modelling of tokamak plasmas for magnetic control validation is based on the so-called Grad-Shafranov partial differential\linebreak equation~(GS-PDE,~\cite{Shafranov:main}).
Because of the low plasma mass density, inertial effects can be neglected and, as a consequence,
the plasma momentum equilibrium equation becomes~$J \times B
= \nabla p$. This equation can be rewritten, in axial-symmetric geometry with cylindrical coordinates~$(r,\phi, z)$,  as:

\begin{equation}
\begin{aligned}
& \Delta^* \psi=-f \frac{df}{d\psi}-\mu_0 r^2\frac{dp}{d\psi} 
\qquad \quad \;
 \text{in the plasma region}  \\
& \Delta^* \psi=-\mu_0 r j_{ext}(r,z,t) 
\qquad \quad \quad
 \text{in the conductors}  \\
& \Delta^* \psi=0 
\qquad \qquad \qquad \qquad \quad \quad 
 \text{elsewhere} \
  \end{aligned}
\end{equation}

\noindent with boundary conditions:
\begin{equation}
    \psi(r,z,t)=\psi_0(r,z),\, \psi(0,z,t)=0,\, \lim\limits_{r^2+z^2 \to \infty} \psi(r,z,t) = 0, \forall t
\end{equation}

\noindent where $\psi=\psi(r,z)$ is the poloidal flux per radian,~$\mu_0$ is the vacuum magnetic permeability,~$j_{ext}$ is the toroidal current density in the external conductors (both control coils and passive structures),~$p = p(\psi)$ is the kinetic pressure profile, an~$f = f(\psi)$ is the poloidal current function profile, and the~$\Delta^*$ operator is defined as:
\begin{equation}
    \Delta^*=r\frac{\partial }{\partial r} (\frac{1}{\mu_r r}\frac{\partial \psi}{\partial r})+\frac{\partial }{\partial z}(\frac{1}{\mu_r}\frac{\partial \psi}{\partial z})\,.
\end{equation}

Solutions of the~GS-PDEs can be numerically found by
means of numerical integration techniques such as Finite Element Methods~(FEMs), provided that the plasma boundary can be determined, the toroidal current densities in the~PF coils and the total plasma current are known, functions~$p(\psi)$ and~$f(\psi)$ are defined.~$j_{ext}$ can be expressed as a linear combination of the circuit currents, the time evolution of which is given by a circuit equation in the form:

\begin{equation}
    \dot{\Psi}+RI=V
\end{equation}
\noindent Hence, it can be shown that:
\begin{equation}
    j_{ext}=-\frac{\sigma}{r} \dot\psi+\frac{\sigma}{2\pi r}u
    \label{equation:voltage}
\end{equation}

\noindent where~$u$ is the voltage applied to the coils (zero for the passive structure) and~$\sigma$ is the electric conductivity. Equation~(\ref{equation:voltage}) must be integrated over the conductor regions.

Once the~$\psi$ map evolution is known, it is possible to compute other variables of interest for control as plasma current~$I_p$, plasma position and plasma shape. Shape is then described by means of plasma-wall distances at given points (plasma-wall gaps)~\cite{Beghi:CSM} which are usually controlled.

In practice, the~GS-PDE is solved by using numerical solvers, and in  
the present work the CREATE-NL+ nonlinear magnetic equilibrium code~\cite{Albanese:CREATENLnew} is used. This code is exploited in Section~\ref{section:SimResults} to assess the performance of the proposed~ES-based~VS system for the~ITER case.

\medskip
The difficulty of using nonlinear~FEM models for control design purposes makes a linearization procedure of the plasma response necessary. Following the procedure described in~\cite{albanese:CREATEL}, we can finally obtain a linear plasma-circuit dynamics in the form:

\begin{equation} \label{LdotI+RI}
\begin{aligned}
&  L\delta \dot I + R\delta I = \delta V+ L_E\delta \dot w \\
&  \delta y = C \delta I + F \delta w
  \end{aligned}
\end{equation}

\noindent where~$\delta V$ is a vector containing the voltages applied to the circuits (zero for the passive structures) and~$\delta I = [\delta I_A^T ,\delta I_E^T ,\delta I_p ]^T$ is the vector of~PF, passive, and plasma currents respectively; ~$R$ is the circuit resistance matrix and~$L$ is the matrix of self and mutual inductances between the plasma, the coils and the equivalent circuits modelling the passive structures;~$\delta w$ is a vector of parameters describing plasma current density profile, typically assumed as external disturbances (the so called poloidal beta~$\beta_{pl}$ and internal inductance~$l_i$ are often chosen);~$\delta y$ is a vector of outputs and~$C$
and~$F$ are suitable output matrices. All the quantities in which the symbol $\delta$ appears are intended to be variations with respect to the equilibrium value (i.e. the nominal conditions around which the linearization is made).

Assuming:~$$x=I-L^{-1}L_Ew$$ equation~(\ref{LdotI+RI}) can be rewritten in the state space form as:
\begin{equation} \label{Linearized}
\begin{aligned}
&  \delta \dot x = A\delta x+B\,{\rm sat}(\delta V)+E\delta w \\
&  \delta y = \hat C \delta x+F\delta w
  \end{aligned}
\end{equation}
\noindent with clear meaning of symbols and matrices. The limitation of voltage has been accounted for by introducing a saturation function, namely sat~($\cdot$).

\section{THE PLASMA MAGNETIC CONTROL ARCHITECTURE}\label{section:MagneticControlProblem}
In this section the plasma magnetic control problem in a tokamak is briefly discussed, with particular attention to the vertical stabilization problem. 

\begin{figure}[!thb]
\begin{center}
\includegraphics[width=12cm]{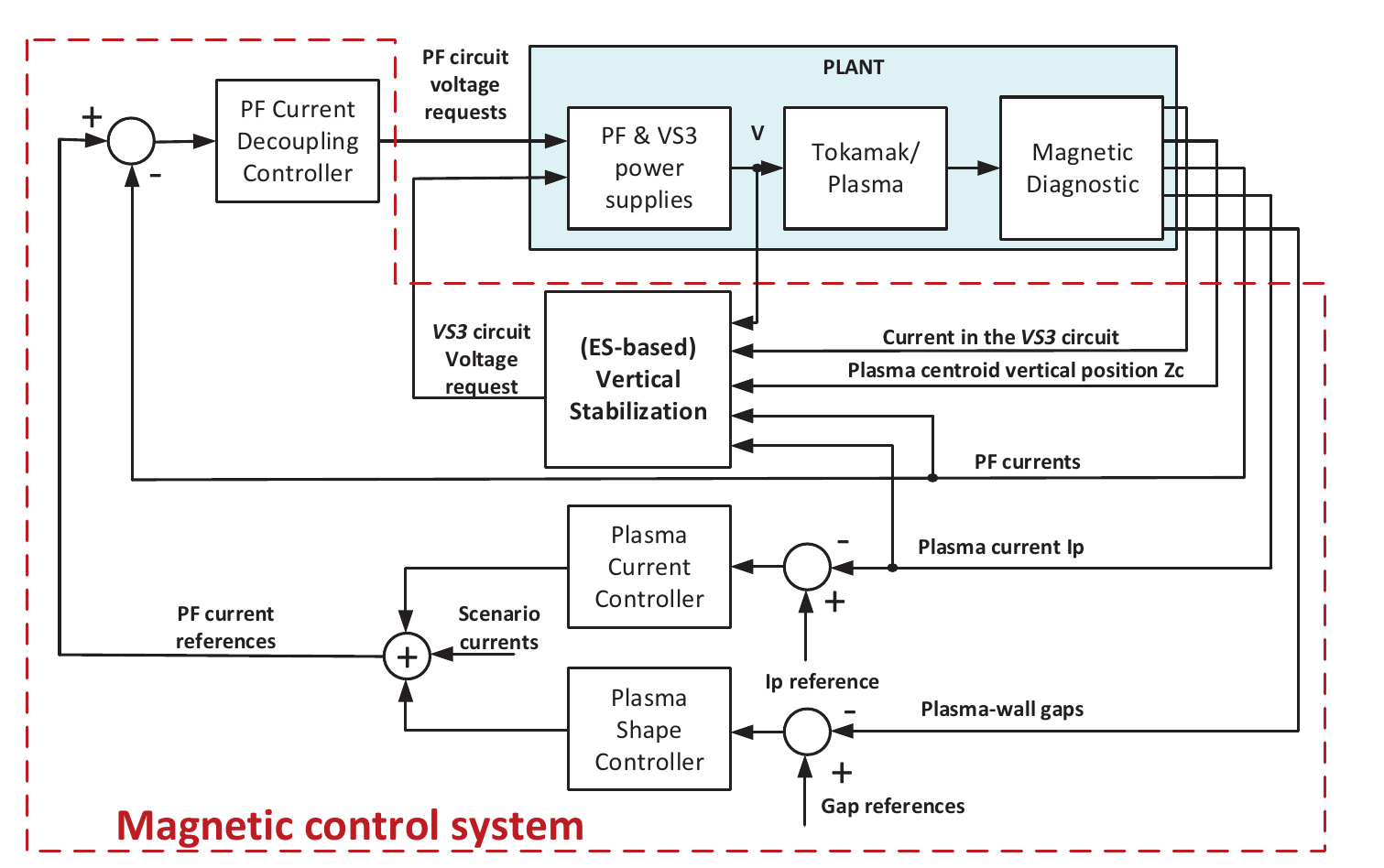}
\caption{Block diagram of a typical~ITER-like architecture for plasma magnetic control in tokamaks~\cite{ambrosino2015design,cinque2019management}. The~ES-based~VS algorithm proposed in Section~\ref{section:VS_ES} is meant to be deployed in the \emph{Vertical Stabilization} block reported in this diagram.}\label{figure:Architecture}
\end{center}
\end{figure}

The simplified block diagram of a possible magnetic control architecture is reported in Figure~\ref{figure:Architecture}. This architecture, other than being broadly adopted in many operating tokamaks, such as~JET~\cite{Sartori:JET} and~EAST~\cite{albanese2017iter}, is also the one currently considered for~ITER~\cite{ambrosino2015design,cinque2019management}. 

\medskip
As already mentioned in Section~\ref{section:Introduction}, the confinement of the hot plasma in a tokamak device is achieved by means of magnetic fields through the pulse phases defining the so-called plasma scenario. In particular, the magnetic field produced by the~PF coils is needed from the start of the discharge to achieve the conditions for plasma formation inside the vacuum chamber (the so called breakdown and burnthrough phases~\cite{jackson2011control}). Soon after plasma formation, the currents flowing in the~PF coils need to be controlled in order to increase the plasma current during the ramp-up phase, to keep it almost constant during the so-called flat-top, and then to ramp it down during the final phase of the discharge. In addition to the control of the plasma current, also the plasma boundary and position need to be controlled to achieve the desired experimental objectives. Moreover, in the case of vertically elongated plasmas, as in the case of~ITER, the active control of the current in some of the~PF coils is mandatory in order to generate the radial field needed to vertically stabilize the plasma column~\cite{lazarus1990control,walker2009feedback}.

\medskip
The main components of the plasma magnetic control system shown in Figure~\ref{figure:Architecture} are briefly described hereafter\footnote{For more details on the control algorithms implemented by the various blocks shown in~Figure~\ref{figure:Architecture}, the interested reader can refer to~\cite{AriolaPironti:Springer} or~\cite{de2019plasma}.}.

\begin{itemize}
\item The~\textbf{PF Current~(PFC) Decoupling Controller}, this block acts as the inner control loop of a nested architecture that includes also the plasma current and shape controllers. By generating the required voltages to be applied to the superconductive coils, this block tracks the PF current references, which are a sum of the \emph{scenario} (i.e.,~the nominal) currents and the corrections requested by the outer loops to track the desired plasma shape and current;
\item The \textbf{Plasma Current Controller}, which tracks the plasma current reference by sending the correspondent requests to the~PFC Decoupling Controller;
\item The \textbf{Plasma Shape Controller}, which controls the shape of the last closed flux surface within the vacuum chamber by tracking a set of plasma shape descriptors; this block also generates requests for the~PFC Decoupling Controller.
\item The \textbf{Vertical Stabilization} (VS) system, which is in charge of vertically stabilizing the plasma column.
More details about this block are given in the next section. \end{itemize}

\subsection{THE VERTICAL STABILIZATION PROBLEM}\label{subsection:VSproblem}
High performance plasmas have a diverted shape~(i.e., with an \emph{active}~X-point in the vacuum chamber) with an elongated poloidal cross-section, as the one shown in Figure~\ref{figure:ITER_pcs}. Indeed, elongated plasmas provide considerable advantages on energy confinement and achievable pressures. The price to be paid to improve the fusion performance is that such elongated plasmas are vertically unstable (a simple description of such instability can be found in~\cite{de2019plasma}). 

\medskip

The plasma vertical instability reveals itself in the linearized model of the plasma behaviour~(\ref{Linearized}) by the presence of an unstable eigenvalue. 
Thanks to the presence of the conducting structures that surround the plasma, the instability characteristic time is brought to a scale that can be controllable via active stabilization circuits. 

It follows that the Vertical Stabilization block in Figure~\ref{figure:Architecture} is an essential component of the magnetic control algorithm to run tokamak discharges with an elongated plasma. The stability must be guaranteed in the presence of uncertainties and time varying behaviour of the plasma along the scenario, and good performance of the overall plasma magnetic control system should be guaranteed in the presence of 
disturbances, such as Edge Localized Modes~(ELMs) or Minor Disruptions (MDs)~\cite{corona2019plasma}, and other fast disturbances modeled as Vertical Displacement Events~(VDEs,~\cite{CST:2010})\footnote{A~VDE is a sudden and uncontrolled displacement of the plasma centroid in the vertical direction. This is often used as a benchmark for the best performance achievable by a Vertical Stabilization system}. In practically all the existing tokamaks, the~VS system drives a combination of currents in a set of dedicated~PF circuits that produces a mainly radial magnetic field which is needed to apply the vertical force used to stop the plasma column. In~ITER such dedicated circuit is the so called~\emph{VS3} circuit (see again Figure~\ref{figure:ITER_pcs}), which is made by one pair of coils fed in anti-series\footnote{Two coils are said to be connected in anti-series if they are connected together with the winding, and hence the flowing current, in opposite direction.}.

VS control algorithms with a simple structure and few control parameters are usually preferred on existing machines~\cite{albanese2017iter}. Indeed, a simple structure enables the deployment of effective adaptive algorithms, aimed at robust operations under various scenarios~\cite{Neto:CEP2012}. However, such adaptive algorithms are not always straightforward to design, either because they require reliable models, which are not necessarily available, or because their tuning requires a considerable effort in terms of time, as well as a considerable experience on the specific machine. Therefore, in the next section, we propose an~ES-based~VS system that aims at achieving the requested robustness without the need of a detailed model, thanks to the model-agnostic nature of the ES algorithm.

\section{ES-BASED PLASMA VERTICAL STABILIZATION}\label{section:VS_ES}
The~ES-based architecture for the~VS of tokamak plasmas proposed in this work is based on the approach described in~\cite{scheinker2017model} to stabilize an unknown, unstable system. However, in order to apply it to solve the~VS problem, an estimation of the state motion along the unstable mode is needed; one possibility to obtain such estimate is to use an observer such as a Kalman filter.

The~ES method presented in~\cite{scheinker2017model} aims at achieving stabilization via minimization of a candidate Lyapunov function of the unstable system, assuming that the state can be measured or estimated.
In particular, it has been shown that given the nonlinear system affine in control
\begin{equation*}\label{equation:nonlinearsystemaffineincontrol}
\dot{x}(t) = f(x,t)+g(x,t)u(t)\,,
\end{equation*}
it is possible to employ a nonlinear time-varying control law in the form
\begin{equation}\label{equation:EScontrollaw}
u(t) = \alpha \sqrt{\omega} cos(\omega t) - k \sqrt{\omega} sin(\omega t)V(x)\,,
\end{equation}
where~$V(x)$ is a Lyapunov function, used to stabilize the associated Lie-bracket average system

\begin{equation}\label{equation:ESaveragesystem}
\dot{\bar{x}}(t) = f(\bar{x},t)-k \alpha\,g(\bar{x},t) g^{T}(\bar{x},t)\left(\frac{\delta V(\bar{x})}{\delta \bar{x}} \right)^{T}\,,
\end{equation}
In fact, from eq.~\eqref{equation:ESaveragesystem} it can be seen how a choice of a sufficiently high positive gain~$k \alpha$ makes the gradient term dominant and the average system asymptotically stabilized.

Moreover, it can be shown, via averaging arguments, that the trajectories of system~\eqref{equation:nonlinearsystemaffineincontrol} under the control input~\eqref{equation:EScontrollaw} can be kept arbitrarily close to those of \eqref{equation:ESaveragesystem},
%
%
provided that the frequency~$\omega$ is chosen high enough. This guarantees that all the trajectories of the original system are confined to a neighborhood of the averaged ones, making the system semi-globally practically stabilized (more details can be found in~\cite{scheinker2017model},~\cite{871771} and~\cite{TEEL1999329}).

Although the stabilization via~ES does not require the knowledge of the system, it requires that the value of the function~$V(\cdot)$ is known, which means tha system's state must be accessible.
For a tokamak plasma, like the~ITER one, only a subset of the state of the associated linearized system~(\ref{Linearized}) can be measured or readly estimated in real-time with a static combination of measurements, which in this case consists in the~PF currents~$I_{A}$ and the plasma current~$I_{p}$, while the eddy currents would require a dynamic estimator. 
On the other hand, as a matter of fact, is not possible to define a candidate Lyapunov function disregarding the eddy currents which play a fundamental role in the dynamics of the vertical instability.

Therefore, in this work, we propose to use a candidate Lyapunov function based only on the estimation of the state dynamics along the unstable mode of the linearized model~(\ref{Linearized}). 
This estimation is achieved by means of a Kalman filter. Although such a filter requires the knowledge of a model, in Section\ref{section:SimResults} it is shown that the proposed architecture can cope with relevant model uncertainties, since it anyhow exploits the \emph{model-agnostic} nature of the~ES algorithm~\eqref{equation:EScontrollaw}. 

\begin{figure}[!thb]
\begin{center}
\includegraphics[width=8cm]{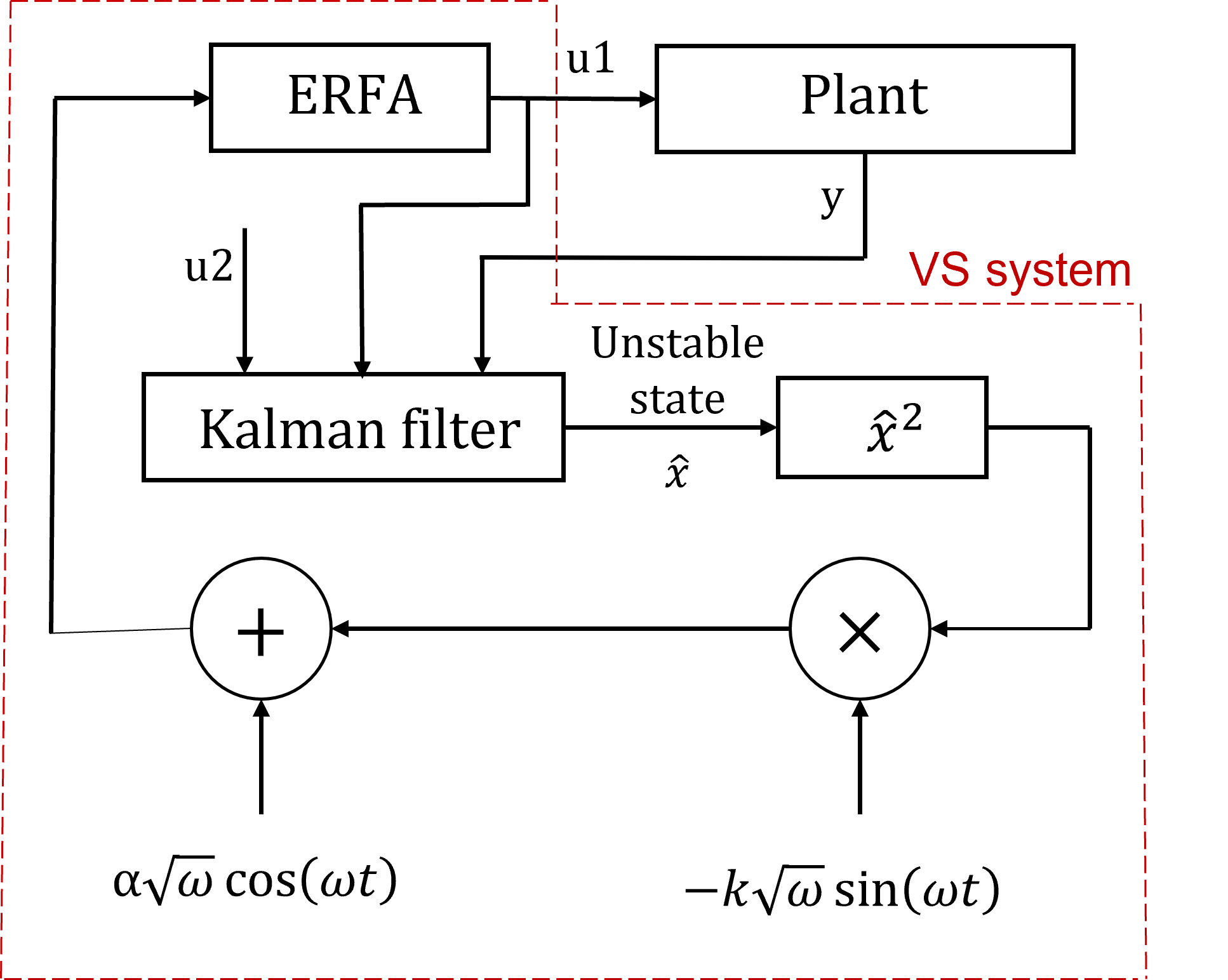}
\caption{The proposed~VS system based on the~ES stabilization algorithm and the switching power supply.} \label{figure:ES_kalman_ERFA}
\end{center}
\end{figure}

\medskip
 A scheme of the~ES-based~VS architecture implemented for~ITER is shown in Figure~\ref{figure:ES_kalman_ERFA}. The control output of the stabilization system~$u_{1}$, is the voltage request to the~ITER~VS circuit~$VS3$. The other input to the plant, $u_2$, is a vector containing the voltages applied to the superconductive~PF circuits by the PF Current Decoupling Controller (see Figure~\ref{figure:Architecture}). For the proposed application, the considered plant output vector~$y$ is defined as 
 \[
y = \begin{pmatrix}\delta I^T_{A} & \delta I_{p} & \delta Z_{c}\end{pmatrix}^T\,,
\]
 
i.e. it contains the variations, with respect to the equilibrium value, of the currents in the active circuits~$\delta I_A$ (both the superconductive~PF and the~$VS3$ circuits), of the plasma current~$\delta I_p$ and of the vertical position of the plasma centroid~$\delta Z_c$. It is worth to observe that, in a real tokamak, the active currents~$I_A$ can be directly measured, while the plasma current and the centroid vertical position are usually obtained as a linear combination of the available magnetic field measurements. On the other hand, for what concerns the plant inputs, the equilibrium voltages are all equal to zero, therefore the variations of~$u_1$ and~$u_2$ coincide with their actual values\footnote{In the first approximation, the plasma is assumed to be superconductive, so there is no need to ramp the PF~currents in order to compensate for the ohmic drop.}.

 
The Kalman filter receives as input~$u_1$, $u_{2}$ and~$y$ and provides an estimation of the dynamics along the unstable mode~$\hat{x}$. 
It has been designed assuming high confidence in the measurements, which is reflected in the choice of almost negligible covariance matrices. The estimation of~$\hat{x}$ returned by the Kalman filter is used to compute the candidate Lyapunov function~$V(\hat{x})=\hat{x}^2$ to be minimized by the~ES control algorithm.

The tuning of the control gains~$k$ and~$\alpha$ in~\eqref{equation:EScontrollaw} can be carried out with a trial and error procedure by means of numerical simulations. However, a first guess for the product~$k \alpha$ has been obtained by considering the first order reduced model that links the voltage applied to the vertical stabilization circuit to the unstable state (i.e. the unstable dynamics alone). Indeed, when such first order reduced model is considered, from~\eqref{equation:ESaveragesystem} it readily follows that the closed loop average system is equal to~$\dot{\bar{x}} = \left(a_{red}-k \alpha\,b_{red} b_{red}^{T}\right)\bar{x}$; 
therefore, if~$a_{red}>0$ is the unstable eigenvalue of the reduced system, the corresponding average system is stable if the product~$k\alpha$ is sufficiently high.


In order for the averaging arguments that lead to~\eqref{equation:ESaveragesystem} to be valid, the frequency~$\omega$ must be chosen ``high enough"; for this reason, as it is commonly found in averaging analyses, the resulting system exhibits an intrinsic time-scale separation.
In fact, in~\cite{MED2021} a first attempt to apply this technique to the tokamak vertical stabilization problem was proposed, where a linear power amplifier was employed. As a result, the switching frequency~$\omega$ was limited due to the bandwidth of the power supply.
%
%
Conversely, the power supply model that we consider in this paper is a switching one, similar to the one used for the~JET VS system, which is based on the integrated gate commuted thyristors~\cite{toigo2007conceptual}. The availability of a faster actuator enables the choice of  a higher switching frequency~$\omega$ for the mixing and dithering terms in~\eqref{equation:EScontrollaw}, leading to an improvement of the performance with respect to what preliminarly presented in~\cite{MED2021}.

\begin{figure}[!thb]
\begin{center}
\includegraphics[width=7cm]{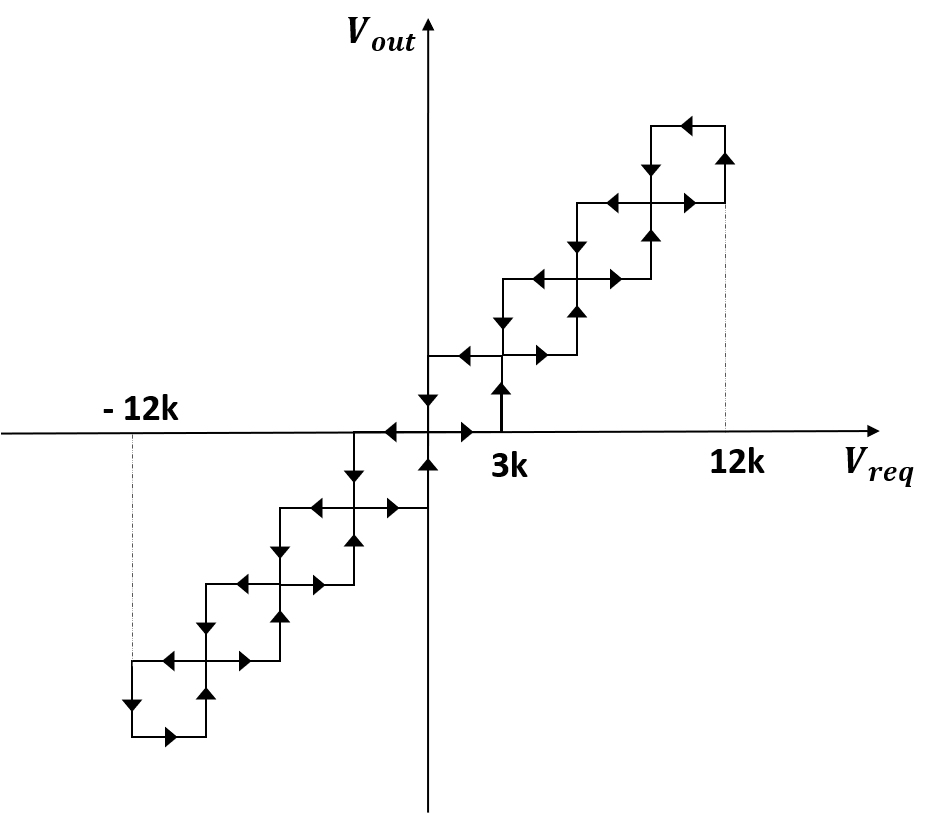}
\caption{Characteristic of the switching power supply. }\label{figure:ERFA_characteristic}
\end{center}
\end{figure}

The characteristic of this kind of power supply, which exhibits a multi-level hysteresis, is reported in Figure~\ref{figure:ERFA_characteristic}. As for the parameters of the power supply, we considered the same of the one currently used at JET, which has a maximum voltage of~$12$~kV, with steps of~$3$~kV~(see~Table~\ref{table:ERFAparameters}). 


\begin{table}[h!t]
\centering
\begin{tabular}{ l|c c c c } 
 \hline
 Parameters & ERFA  \\ [1ex]
 \hline 
 Max output voltage & $\pm 12$~kV dc \\[1ex]
 Max output current & $\pm 5$~kA dc  \\[1ex]
 Max output voltage step & $\pm 3$ kV \\[1ex]
 Time for full $\pm$ voltage excursion & $\leq 100 \mu$ s \\[1ex]
 Max switching frequency & $1$ kHz \\[1ex]
 \hline
 \end{tabular}
  \caption{Main parameters of the fast switching power supply.}
\label{table:ERFAparameters}
\end{table}

The power supply considered in this work has been simulated taking into account the maximum output voltage and voltage steps reported in Table~\ref{table:ERFAparameters}, with an internal delay of~$200~\mu s$ . 


\section{SIMULATIONS OF THE PROPOSED CONTROL ALGORITHM}\label{section:SimResults}

The proposed ES-based VS has been tested by both linear and nonlinear simulations, with the aim of proving the validity of the approach and to assess its robustness. 

As already pointed out in Section~\ref{section:Introduction}, the control architecture exploited for the simulations includes the overall~ITER magnetic control system, whose scheme is reported in Figure~\ref{figure:Architecture}; therefore the interaction of the proposed ES-based VS with the plasma current and shape control systems is also taken into account. 
In particular, the shape control algorithm adopted is the so-called eXtreme Shape Controller, which in this case controls~$29$ plasma-wall distances (the \emph{gaps} shown in Figure~\ref{figure:Gapsnl}), with a settling time of about~$10~s$. In order to control a number of plasma shape descriptors, i.e. the~29 gaps, which is greater than the number of available actuators (i.e., the~11 currents in the superconductive coils), the~XSC design is based on a Singular Value Decomposition of the static relationship between the control inputs and outputs. In this way, it is possible to minimize in least mean square sense the control error at steady-state. For more details about the XSC the interested reader can refer to~\cite{Ariola:XSC,xscJET}.

Moreover, the simulation have been performed considering a set of operational scenarios for the~ITER tokamak. The considered scenarios refer to the counteraction of relevant disturbances that can occur during plasma operation.
In particular, the following cases were considered:
\begin{itemize}
 \item the rejection of a~VDE of~5~cm;
 \item the response to a~MD, 

\end{itemize}

A VDE is an instantaneous motion of the system's state along the unstable mode, scaled so as to produce a prescribed vertical displacement of the plasma centroid. The controller must be able to counteract this displacement, bringing back the centroid to the reference position (i.e. zero displacement with respect to the equilibrium value).
A MD represents instead a lost of a fraction of the plasma thermal energy, due to the uncontrolled growth of some plasma instability. For this application, a~MD can be modelled as an instantaneous drop of $0.1$, from the nominal values, of the disturbance parameters~$\beta_p$ and~$l_i$.  

For all the considered scenarios, the same configuration of the~VS system has been used. This means that, in all simulations, the same Kalman filter was taken into account, as well as the same ES parameters. The value of the ES control parameter in the control law~\eqref{equation:EScontrollaw} is reported in Table~\ref{table:ESparameters}, while the parameters of the switching power supply are the shown in Table~\ref{table:ERFAparameters}.

\begin{table}[!h]
\centering
\begin{tabular}{c c c } 
 \hline 
 $k$ & $\alpha$ & $\omega$ \\ [1ex]
 \hline
  $2.7\cdot 10^{-3}$ & $1$ & $250\cdot 2\pi$~rad/s \\[1ex]
 \hline
 \end{tabular}
  \caption{Control parameters for the proposed model-free~VS system based on the~ES control law~\eqref{equation:EScontrollaw}.} 
\label{table:ESparameters}
\end{table}

\begin{figure}[h]
\begin{center}
\includegraphics[width=0.11\linewidth, trim= 60 20 60 20]{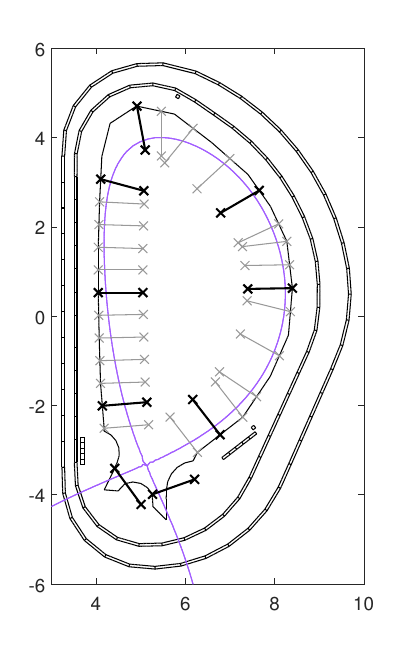}
\caption{Gaps used for the shape control. The gaps shown in black are the ones whose behaviour is reported in Figs.~\ref{figure:gaps}-\ref{figure:Gapsnl}.}
\label{figure:gapsposition}
\end{center}
\end{figure}

\subsection{LINEAR SIMULATION VALIDATION}\label{subsection:LinearSim}

The linear simulations have been carried out with the~VS system architecture presented in Section~\ref{section:VS_ES}. The aim was to prove that the designed ES-based approach can stabilize a broad family of different plasma models, although the embedded Kalman filter is always the same.

The considered family of plasma models consists of~24 different plasma equilibria, all at a plasma current of~$15 MA$, generated so as to cover the interval
\[
(l_i,\beta_p)~\in~\left[ 0.8,~1.3\right]\times\left[0.1,~1\right]\,,
\]
with two different plasma shapes, characterized by two slightly different elongations~$\kappa = 1.81, 1.76$, respectively\footnote{The elongation is defined as~$\kappa=\dfrac{b}{a}$, where~$b$ is the plasma height and~$a$ is the plasma minor radius~\cite[Ch.~5]{Freidberg}. Plasma growth rate increases with the elongation}.

The unique Kalman filter adopted for linear simulations was obtained considering a reduced linearized model, of order~25, for the equilibrium characterized by~$l_i = 1.3$,~$\beta_p = 1 $ and a growth rate~$\gamma = 7.6~s^{-1}$. The operational scenario considered for the linear simulations is a rejection of a~VDE of~5~cm.

\begin{figure}[h]
\begin{center}
\includegraphics[width=0.7\textwidth, trim= 70 20 70 20]{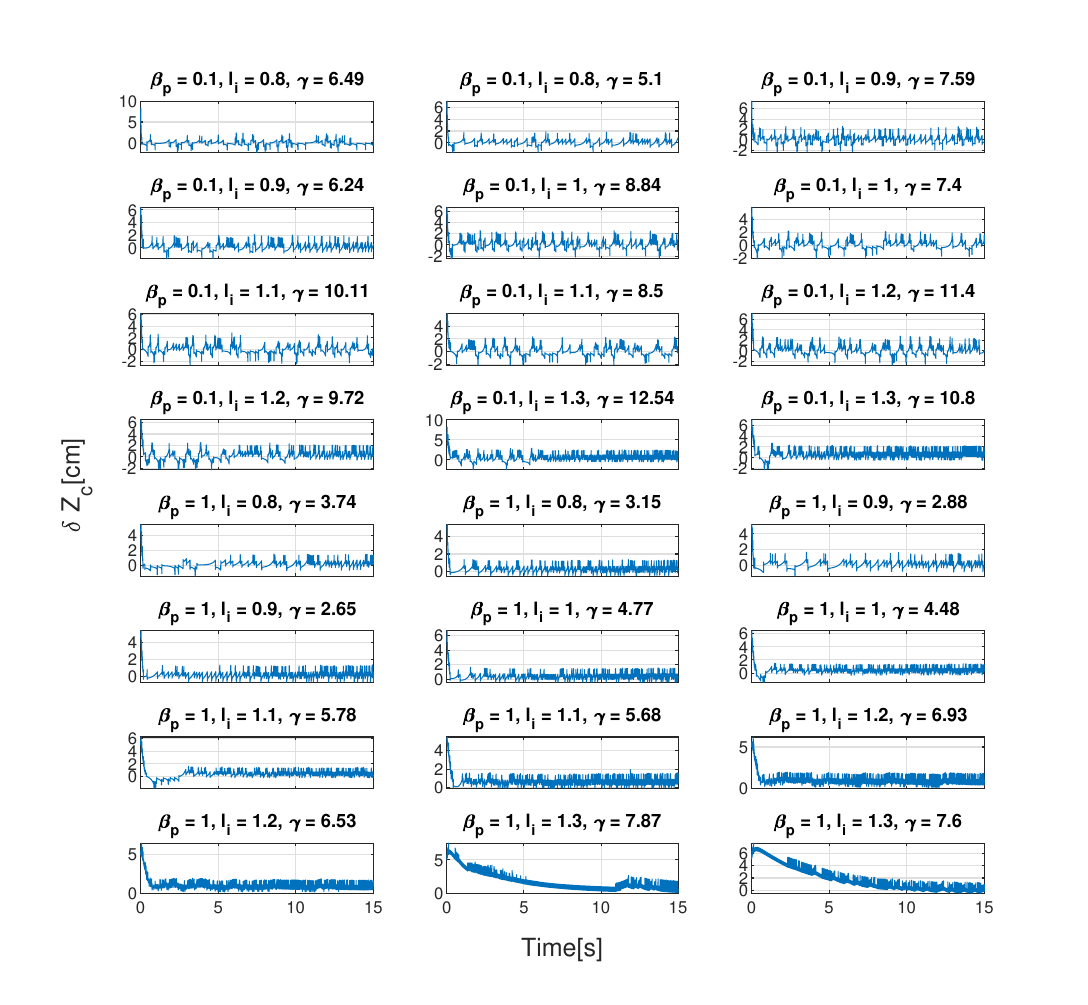}
\caption{Response to a~VDE of~5~cm for the considered family of different plasma models in terms of the displacement from the equilibrium value of the plasma vertical position~$Z_{c}$.} 
\label{figure:Zc}
\end{center}
\end{figure}

\begin{figure}[!thb]
\begin{center}
\includegraphics[width=0.7\textwidth, trim= 70 20 70 20]{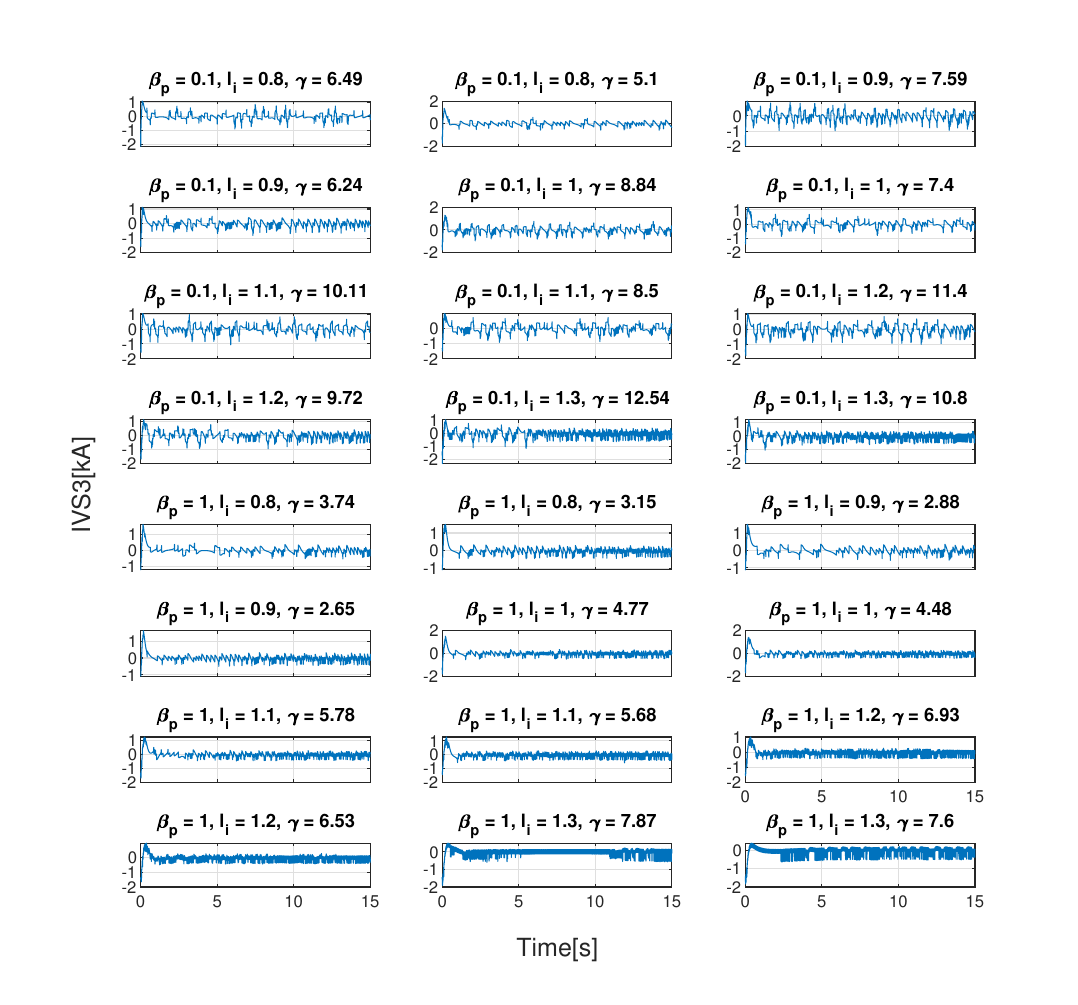}
\caption{Response to a~VDE of~5~cm for the considered family of different plasma. The time behaviour of the currents in the~VS system coils~$IVS3$ have been reported. }\label{figure:IVS3}
\end{center}
\end{figure}

\begin{figure}[!h]
\begin{center}
\includegraphics[width=0.7\textwidth, trim= 70 20 70 20]{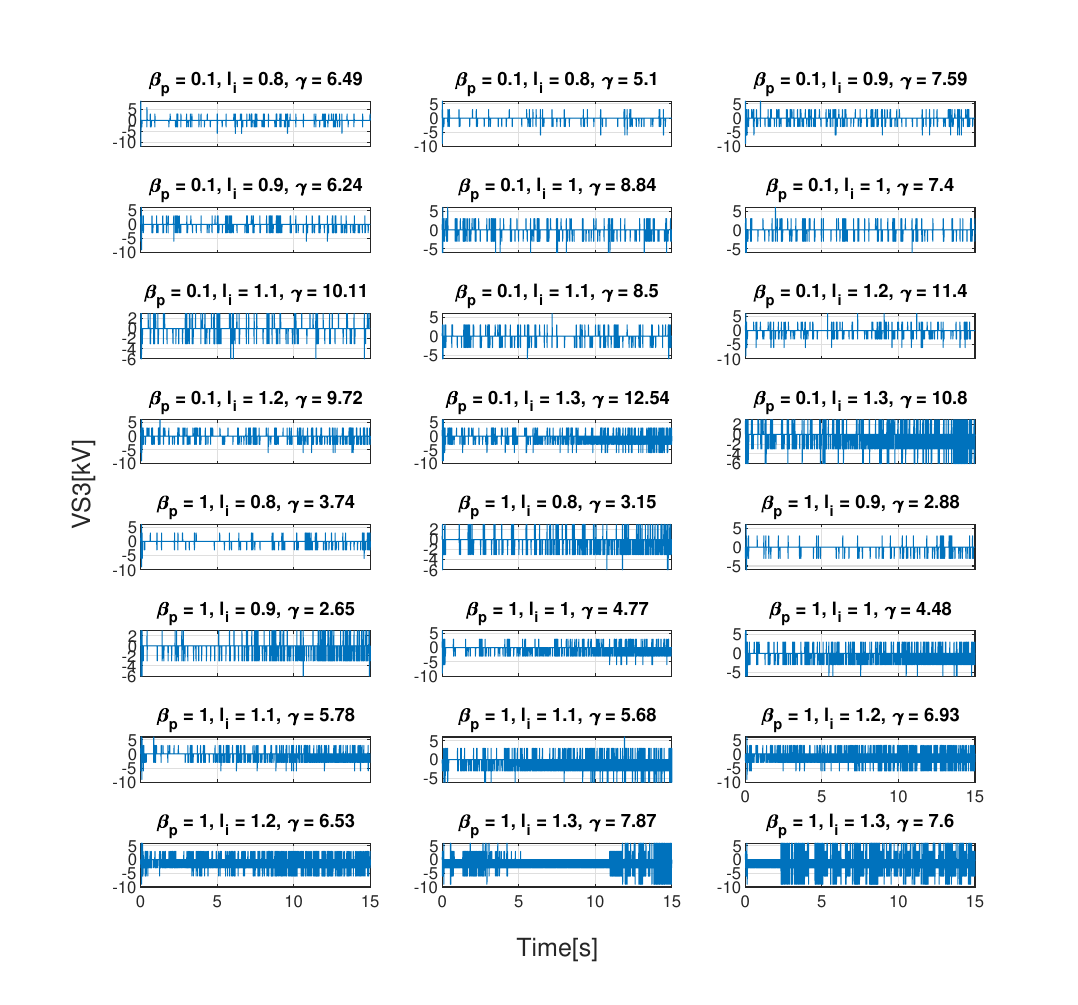}
\caption{Response to a~VDE of~5~cm for the considered family of different plasma model. The figure reports the voltages applied to~VS system,~$VS3$.}
\label{figure:VS3}
\end{center}
\end{figure}

\begin{figure}[!hb]
\begin{center}
\includegraphics[width=0.7\textwidth, trim= 70 20 70 20]{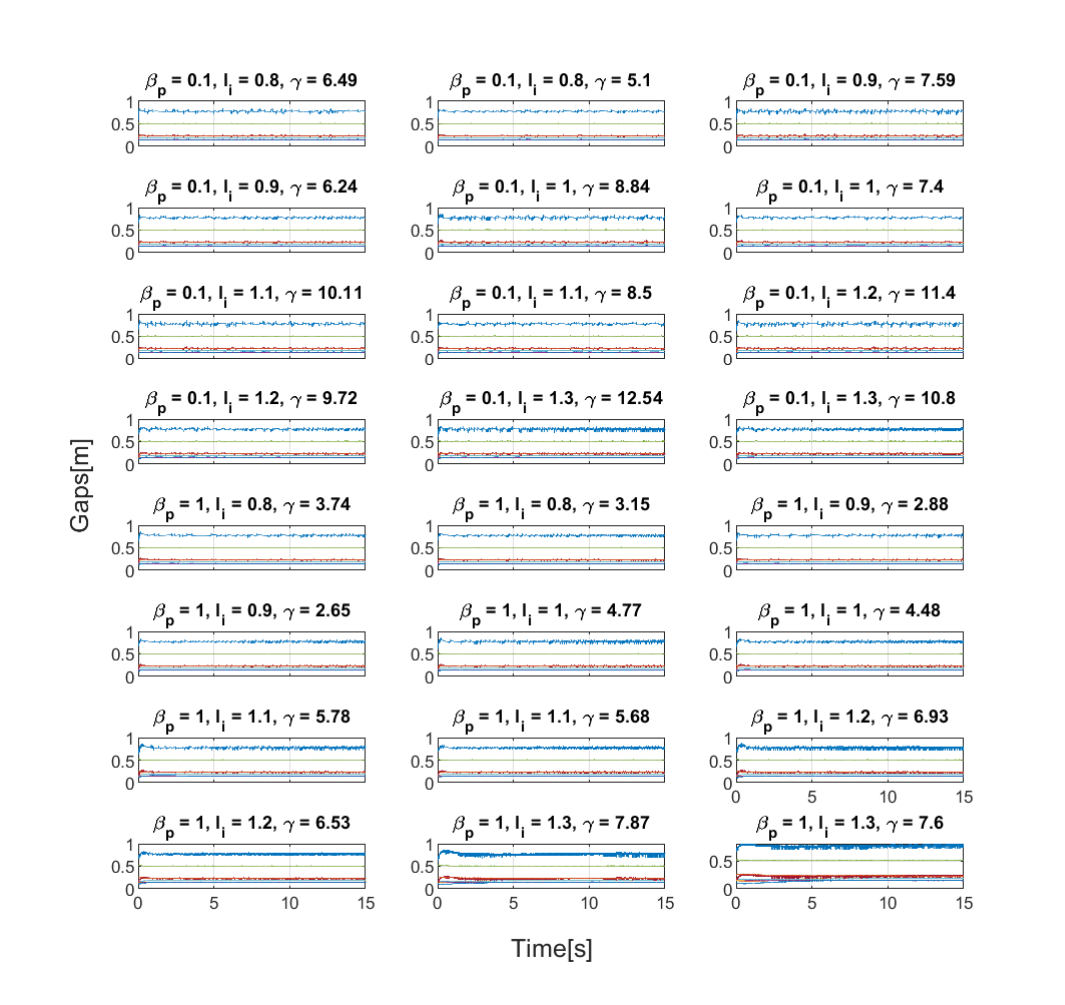}
\caption{Response to a~VDE of~5~cm for the considered family of different plasma model. The time behaviour of the position of some of the main gaps have been reported. }\label{figure:gaps}
\end{center}
\end{figure}

The results of the simulations are shown in Figs.~\ref{figure:Zc}-\ref{figure:gaps}, where the displacement from the equilibrium of the plasma centroid position~$\delta Z_{c}$, the current and voltage in the VS coils, $IVS3$ and $VS3$ respectively, and the value of the main gaps, chosen according to Figure~\ref{figure:gapsposition}, are reported for the family of different plasma models considered. 

The results show that, thanks to the model-agnostic nature of the ES algorithm, the proposed architecture is capable of dealing with relevant model uncertainties. Indeed, all the considered plasma configurations can be stabilized with the use of a single Kalman filter designed to estimate a simplified, reduced order dynamics (as discussed in Section~\ref{section:VS_ES}). This indicates that the proposed VS system can guarantee a satisfactory degree of robustness. 

\medskip
As it can be seen in Figs.~\ref{figure:Zc}-\ref{figure:gaps}, the worst-case maximum plasma vertical displacement is about~$10~cm$, while in most of the cases the maximum $\delta Z_c$ is close to the initial VDE of $5~cm$. Furthermore, the $Z_c$ variation is rejected very rapidly in most of the considered cases, with the exception of the last two configurations shown in~Figure~\ref{figure:Zc} where the settling times reaches a few seconds. Finally, it is worth to observe that in all the considered cases the maximum in-vessel current is in the order of a few~kA.

Plasma current and shape controllers were also included in the simulation scheme, in order to verify that they do not negatively interact with the proposed VS. In fact, it was possible to verify that the plasma current remains practically unchanged (the variations are of the order of a few~kA for a plasma current of~$15~MA$), while the plasma boundary does not touch the vessel during the transient in any of the considered scenarios (Figure~\ref{figure:gaps} shows that the gaps are always positive, i.e. the plasma boundary never collides with the surrounding walls).

\subsection{NONLINEAR SIMULATIONS}\label{nonlinearsimulations}

Apart from the linear simulation discussed in Section~\ref{subsection:LinearSim}, nonlinear numerical simulations have been carried out with the~CREATE-NL+ free boundary evolutionary code, with the aim of validating the proposed VS control approach on~ITER plasma scenarios, in the presence of significant non-linearities and of a more realistic behaviour of the plasma. In the past, the~CREATE-NL+ code has been validated against experimental data coming from several tokamaks, including~JET {}~\cite{Albanese:CREATENLnew}.
%
For the simulations, the three following starting equilibria have been chosen, whose main parameters are summarized in Table~\ref{tab:configs}:
\begin{itemize}
    \item an equilibrium at the end of the ramp up phase~(\emph{Eq \#1} in~table~\ref{tab:configs}).
    \item an equilibrium at the beginning of the flat-top phase~(\emph{Eq \#2} in~table~\ref{tab:configs}).
    \item an equilibrium at the end of the flat-top phase~(\emph{Eq \#3} in~table~\ref{tab:configs}).
\end{itemize}

\begin{table*}[t]
\centering
\begin{tabular}{l | c c c}
\toprule
\textbf{Configuration} & \textbf{Eq \#1} & \textbf{Eq \#2} & \textbf{Eq \#3}  \\
\midrule
\textbf{$I_p~[MA]$} & 14.7 & 15 & 15   \\
\textbf{$\beta_p$} & 0.08 & 0.66 & 0.81 \\
\textbf{$l_i$}     & 0.92 & 0.88 & 0.71 \\
\textbf{$\gamma~[s^{-1}]$}  & 9.1  & 4.9  & 2.9  \\
\textbf{}  & 
\includegraphics[height=3cm, trim= 20 10 30 10]{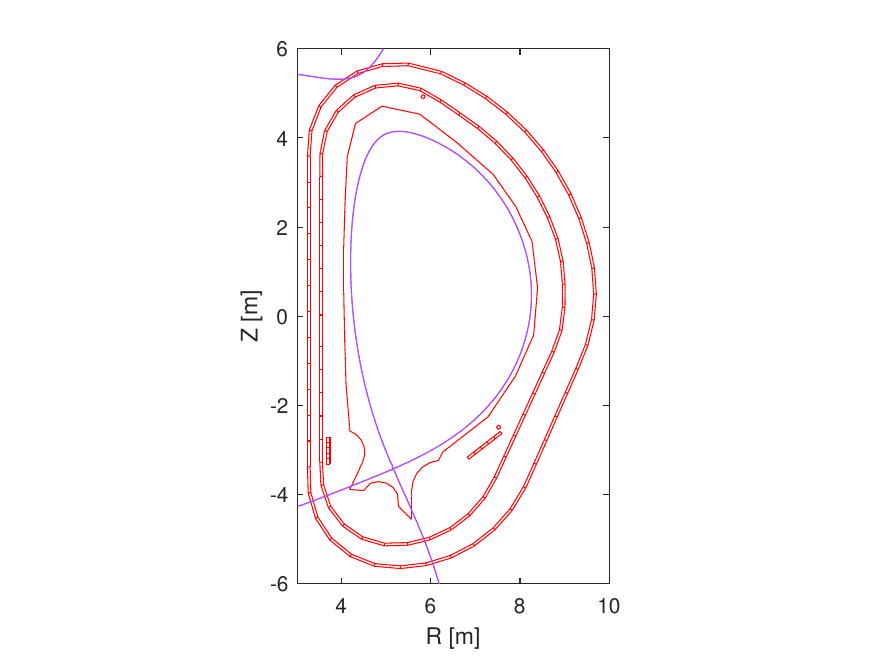}   & 
\includegraphics[height=3cm, trim= 20 10 30 10]{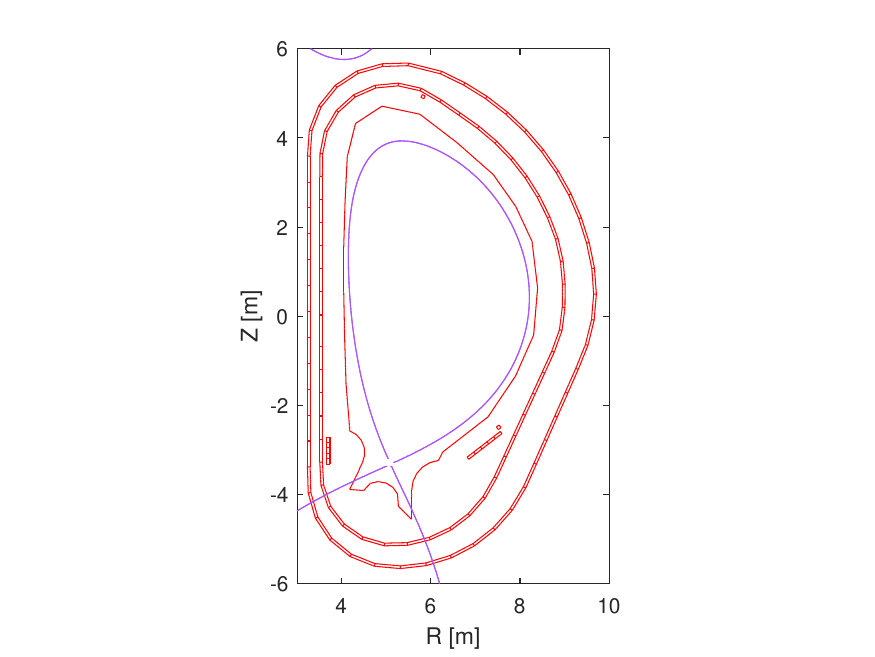}  &
\includegraphics[height=3cm, trim= 20 10 50 10]{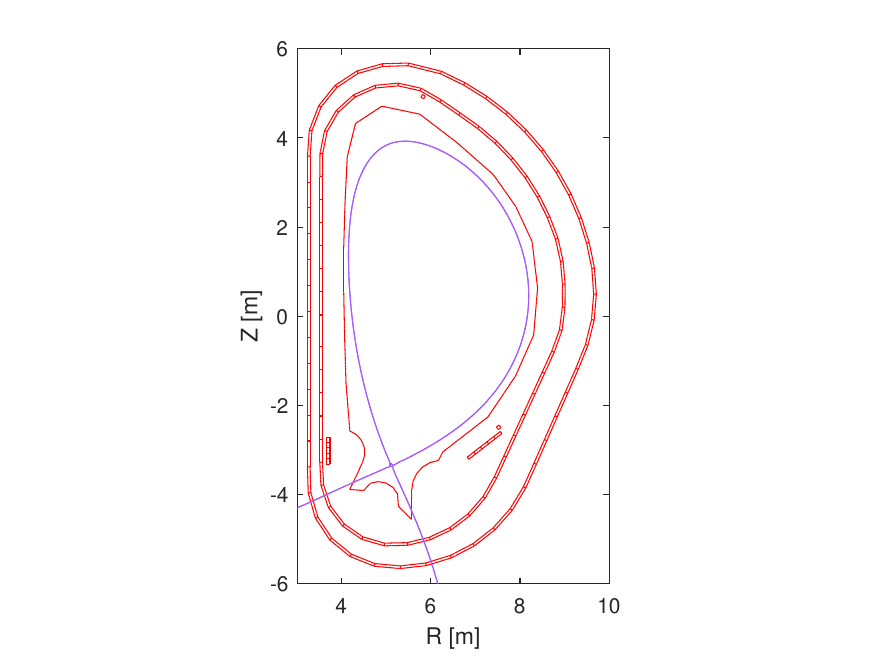}  \\
\bottomrule
\end{tabular}
\caption{ITER equilibria considered for the nonlinear simulations, with the corresponding plasma boundary.}
\label{tab:configs}
\end{table*}

Two different disturbances have been considered:
\begin{itemize}
 \item VDE of~5~cm;
 \item MD ($\Delta \beta_p = - 0.1$, $\Delta l_i = - 0.1$).
\end{itemize}

It is worth to notice that while the disturbance of a  VDE has been applied to every equilibrium, the MD has been considered only for Eq \#2 and Eq \#3 as the MD is a phenomenon arising at higher values of the plasma energy content represented by $\beta_{p}$.


 As in the linear case, the simulations have been carried out including all the blocks of the magnetic control architecture shown in Figure~\ref{figure:Architecture}, taking into account the interaction of the~VS with the plasma current and shape control systems. Moreover, the Kalman filter of the~VS scheme~(Figure~\ref{figure:ES_kalman_ERFA}) has been obtained from the reduced linearized model of~Eq \#2, corresponding to the equilibrium at the beginning of the flat-top phase, while the~ES parameters in the control law~(\ref{equation:EScontrollaw}) are the same used in the linear simulations (see~table~\ref{table:ESparameters}).





\begin{figure}[h!]
\begin{center}
\includegraphics[width=0.7\textwidth, trim= 100 20 90 20]{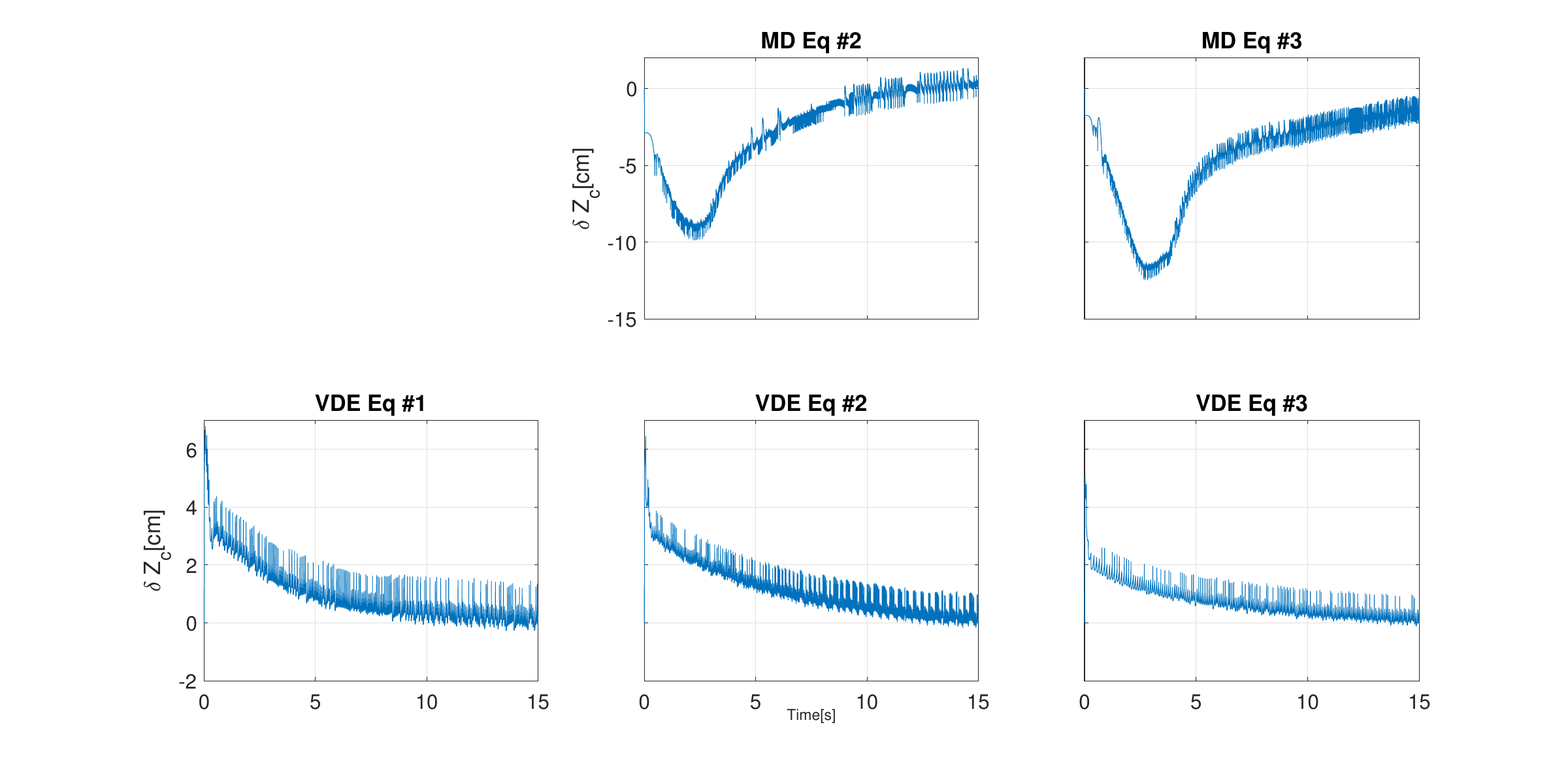}
\caption{Nonlinear response to a MD and a~VDE for the the plasma models in terms of the displacement from the equilibrium values of the plasma vertical position~$Z_c$. }\label{figure:Zcnl}
\end{center}
\end{figure}

\begin{figure}[h!]
\begin{center}
\includegraphics[width=0.7\textwidth, trim= 100 20 90 20]{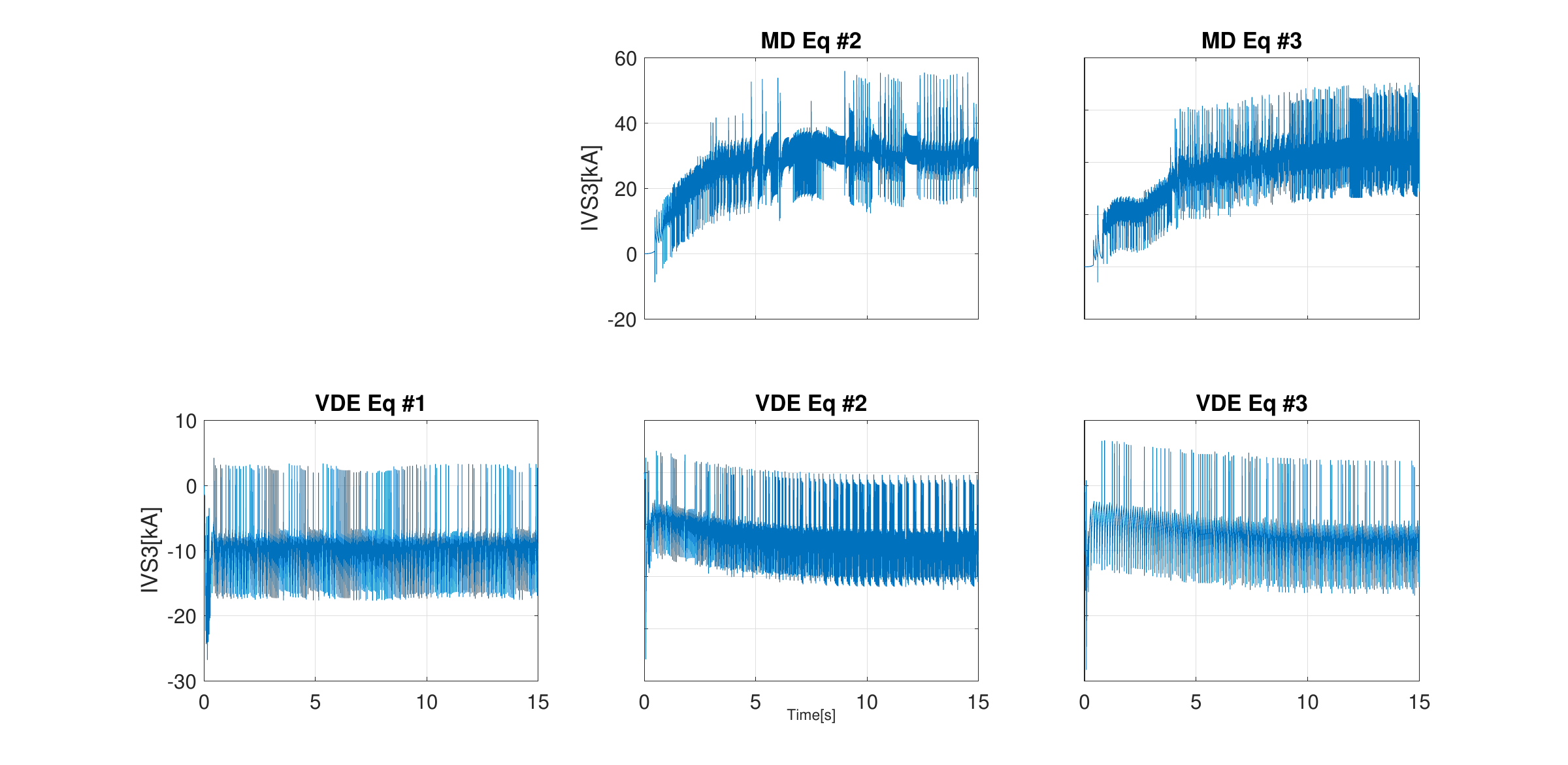}
\caption{Nonlinear response to a~MD and a~VDE for the the plasma models in terms of the current in the~VS system circuit~$IVS3$. }\label{figure:IVSnl}
\end{center}
\end{figure}

\begin{figure}[h!]
\begin{center}
\includegraphics[width=0.87\textwidth, trim= 100 20 90 20]{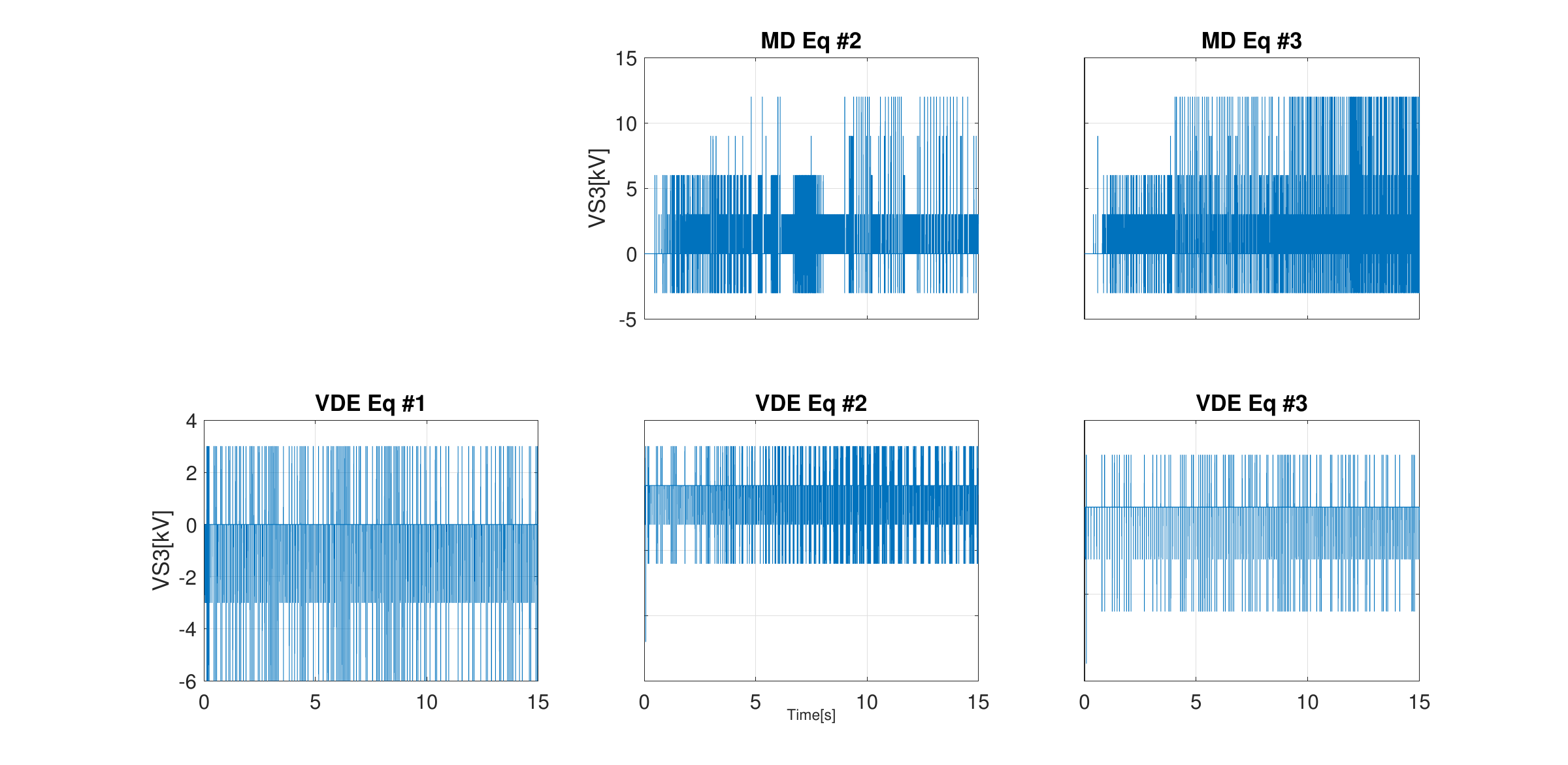}
\caption{Nonlinear response to a~MD and a~VDE for the the plasma models in terms of the voltage applied to the~VS system circuit. }\label{figure:VSnl}
\end{center}
\end{figure}

\begin{figure}[h!]
\begin{center}
\includegraphics[width=0.8\textwidth, trim= 100 20 90 20]{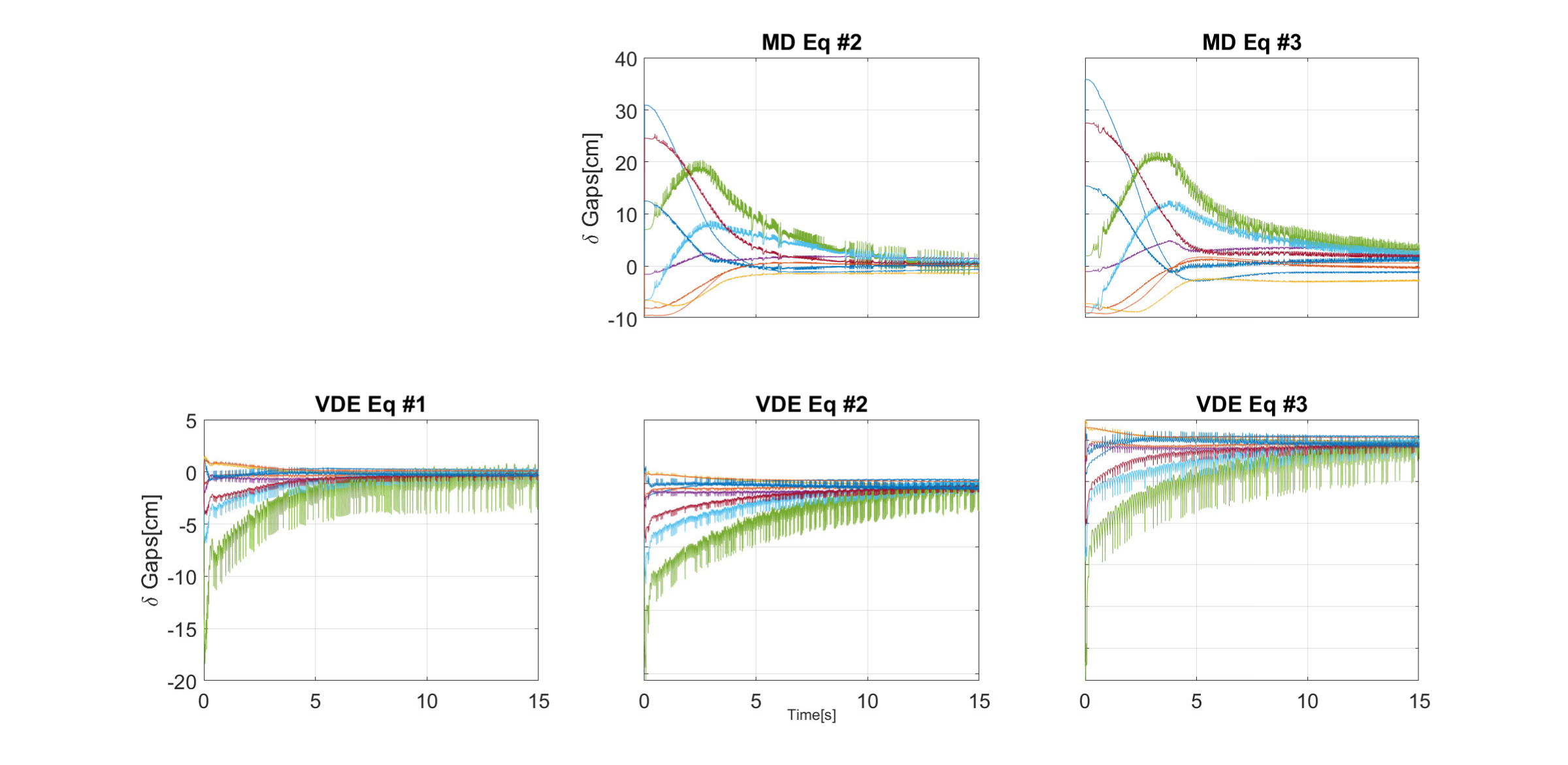}
\caption{Nonlinear response to a~MD and a~VDE for the the plasma models in terms of some of the controlled gaps.}\label{figure:Gapsnl}
\end{center}
\end{figure}

In~Figure~\ref{figure:Zcnl} the displacement of the plasma centroid from the equilibrium position is shown for all the examined cases. It can be noticed that the controllers are able to reject both the considered disturbances starting from the proposed equilibria. The worst case, in terms of~$\delta Z_c$ overshoot, is that of equilibrium~\#3 in the case of a minor disruption.
Moreover, in~Figure~\ref{figure:IVSnl} the current flowing in the stabilization circuit is shown, while~Figure~\ref{figure:VSnl} shows the applied voltage. As expected, the highest current is reached again for~ \emph{Eq~\#3} in the case of a minor disruption. Lastly, in~Figure~\ref{figure:Gapsnl} the gaps evolution is presented, showing how the initial shape is restored after the occurrence of the disturbance.

These simulations underline once again the robustness of the proposed model-free architecture. 

\section*{CONCLUSIONS}
A model-free approach to tackle the plasma VS problem in tokamak devices has been presented in this paper. The proposed~VS architecture consists of a stabilization algorithm based on an~ES-like control law and relies on a single, simplified Kalman filter.

The simulation results show that the proposed VS scheme achieves a satisfactory level of robustness during the overall~flat-top phase of an~ITER discharge. Indeed, the proposed control architecture can practically stabilize the plasma column, by keeping the system state in a bounded set, while counteracting relevant plasma disturbances. Moreover, the \emph{model-agnostic} nature of the~ES algorithm allows to cope with large model uncertainties, due to the different plasma equilibria considered.
In addition, thanks to the inclusion of the plasma current and shape controllers in the simulation scheme, it was possible to verify that the plasma current is not affected by the proposed~ES-based~VS system and the plasma boundary does not touch the first wall in any of the considered scenarios. These results have been validated by means of both linear and nonlinear simulations.


\bibliographystyle{plain}
\bibliography{CEP2021}

\end{document}